\documentclass[12pt]{article}
\usepackage{amsmath}
\usepackage{graphicx}

\newcommand{\be}{\begin{equation}}
\newcommand{\ee}{\end{equation}}
\newcommand{\bea}{\begin{eqnarray}}
\newcommand{\eea}{\end{eqnarray}}

\begin{document}

\bigskip 
\begin{titlepage}

\begin{flushright}
UUITP-17/03\\
hep-th/0309163
\end{flushright}

\vspace{1cm}

\begin{center}
{\Large\bf On Thermalization in de Sitter Space\\}

\end{center}
\vspace{3mm}

\begin{center}

{\large
Ulf H.\ Danielsson{$^1$} and Martin E.\ Olsson{$^2$}} \\

\vspace{5mm}

Institutionen f\"or Teoretisk Fysik, Box 803, SE-751 08
Uppsala, Sweden

\vspace{5mm}

{\tt
{$^1$}ulf@teorfys.uu.se \\
{$^2$}martin.olsson@teorfys.uu.se \\
}

\end{center}
\vspace{5mm}

\begin{center}
{\large \bf Abstract}
\end{center}
\noindent
We discuss thermalization in de Sitter space and argue, from two different points of
view, that the typical time needed for thermalization is of order $R^{3}/l_{pl}^{2}$,
where $R$ is the radius of the de Sitter space in question. This time scale
gives plenty of room for non-thermal deviations to survive during long periods of inflation.
We also speculate in more general terms on the meaning of the time scale for
finite quantum systems inside isolated boxes, and comment on the
relation to the Poincar\'{e} recurrence time.

\vfill
\begin{flushleft}
September 2003
\end{flushleft}
\end{titlepage}
\newpage

\section{Introduction}

\setcounter{equation}{0} \label{sec:intro}

One of very few places where one has realistic hopes of detecting
fundamental physics at or near the Planck scale is in the Cosmological
Microwave Background Radiation (CMBR). The reason for this is that inflation
expands microscopic fluctuations into macroscopic seeds for structure
formation, and thereby possibly imprinting \textit{trans-Planckian} physics
on large scales. For some recent reviews with references see \cite
{Brandenberger:2002sr,kinney2003}. Unfortunately our understanding of
Planckian or stringy physics, and the way inflation is embedded into a
fundamental theory, is not sufficiently well developed to allow for reliable
predictions of what to be expected.

Various models of possible trans-Planckian physics have been proposed
indicating the rough nature of possible effects. In \cite{Ulf:0203} it was
argued that quite independently of the exact mechanism, one can expect a
modulated primordial spectrum with an amplitude linear in $H/\Lambda $. The
main idea was to mimic the unknown physics in terms of a non-standard vacuum
choice at the fundamental scale. This is quite natural since the vacuum is
not unique in an expanding universe. In \cite{Ulf:0205} this was phrased in
terms of a (complex) one parameter family of de Sitter invariant vacua,
called the $\alpha $-vacua, \cite{chernikov,Mottola:ar,Allen:ux,Floreanini:1986tq}.

The introduction of the $\alpha $-vacua has made it possible to analyze the
trans-Planckian proposal in a systematic way. There has been a lively debate
\cite
{Banks:0209,Einhorn:0209,Kaloper:0209,Ulf:0210,Goldstein:0302,Einhorn:0305,
Chung:0305,Collins:0306,Goldstein:0308} as to whether a departure from the 
Bunch-Davies (or \textit{thermal}) vacuum is at all possible and whether field
theoretical arguments put severe constraints on physics that is expected to
be inherently Planckian. In this paper we will not address this issue, but
instead focus on the more general question of to what extent non-thermal
departures from the Bunch-Davies vacuum are possible.

The question we want to discuss is whether, and how long, non-thermal traces
of physics before a possible beginning of inflation can survive and affect
the CMBR. A similar question was asked in \cite{Kaloper:0307} where non-thermal
initial conditions were introduced at a specific time and their
effects on the CMBR analyzed. Technically, as will be explained in the next
section, the analysis is a direct parallel to what was done in \cite
{Ulf:0203} even though the physics is slightly different.

It is reasonable to expect that any non-thermal deviation from the
Bunch-Davies vacuum will experience thermalization due to the Hawking
radiation present during inflation. In this paper we are in particular
interested in the actual thermalization time $\tau $, that is, the maximal
time non-thermal deviations can survive in de Sitter space. We will argue, in
two distinct but possibly related ways, that this time is of order $\tau
\sim R^{3}/l_{p}^{2}$, where $R$ is the horizon radius. This is
substantially longer than the naive expectation of a time scale of the order
$1/T$.

This, one could argue, is generic for quantum systems inside isolated boxes
and does not have anything particularly to do with de Sitter space or
Hawking radiation. With this in mind, we discuss in more general terms the
meaning of the relaxation time mentioned above. In particular we try to
relate it to another typical time scale relevant for finite isolated
systems, namely the Poincar\'{e} recurrence time.

The plan of the paper is as follows. In section 2 we briefly review the
trans-Planckian problem. Particular attention is given to generic signatures
resulting from the assumption that something special happens at a definite
scale, above which new physics is supposed to take over. The ``transverse''
case, where instead something special happens at a definite time during the
rolling of the inflaton field, is also discussed. In section 3 the issue of
thermalization is addressed. As already mentioned, the typical
thermalization time for non-thermal excitations in de Sitter space is argued
to be of order $\tau \sim R^{3}/l_{p}^{2}$. The section ends by attempts to
put this into a more general context. We conclude in section 4.

\section{Signatures of new physics in the CMBR}

\setcounter{equation}{0} \label{sec:sign}

\bigskip

\subsection{Modified initial conditions on a fixed scale}

It has been argued in several works that physics at or beyond the Planck
scale (or string scale) might leave a non negligible imprint in the CMBR.
The basic argument behind the claim is that the rapid expansion of the
universe during inflation magnifies small scale physics and makes it
accessible on large scales. Going back in time the quantum fluctuations
responsible for the structure in the CMBR eventually reaches a linear scale
of the order the Planck scale and one would expect that Planckian physics
determine their form. In fact, in an expanding universe the notion of a
vacuum is not unique. The standard procedure is to ignore the presence of
the Planck scale and trace the quantum state back to the infinite past. In
an accelerated universe (where the scale factor grows faster than the Hubble
radius) the linear scale of the fluctuations can be made arbitrarily smaller
than the Hubble scale and the vacuum becomes unique. In exact de Sitter
space the vacuum is called the Bunch-Davies vacuum and happens to be thermal.

There is clearly something unsatisfactory with this procedure. A more
conservative approach was advocated in \cite{Ulf:0203} where it was argued
that one should stop when the modes reached the Planck scale (or string
scale) and impose initial conditions there. It was shown that a particular
choice of initial conditions (the instantaneous Minkowski vacuum) leads to a
corrected primordial spectrum of the form
\begin{equation}
P(k)=\left( \frac{H}{\overset{\cdot }{\phi }}\right) ^{2}\left( \frac{H}{%
2\pi }\right) ^{2}\left( 1-\frac{H}{\Lambda }\sin \left( \frac{2\Lambda }{H}%
\right) \right) .  \label{eq:pk}
\end{equation}
It was also argued that the linear dependence on $H/\Lambda $ in the
correction term is rather generic. If, furthermore, we are not in exact de
Sitter space but instead experience a slow roll where $H$ is changing with
time (which is translated into scale on the sky) the primordial spectrum
will have a characteristic modulation. See also \cite{Easther:2001fi,
Easther:2001fz,Easther:2002xe}. The precise amplitude and phase of
the modulation depends in a detailed way on the initial conditions chosen,
but the modulation is a generic effect and a typical sign of trans-Planckian
physics. It is amusing to compare the effect with the acoustic peaks which
are due to oscillations \textit{after} the fluctuations have reentered the
horizon. The modifications we are studying have their origin in oscillations
\textit{before} the fluctuations exit.

\bigskip

\subsection{Modified initial conditions at a fixed time}

\bigskip

\begin{figure}[tbh]
\centering
\includegraphics[height=6cm]{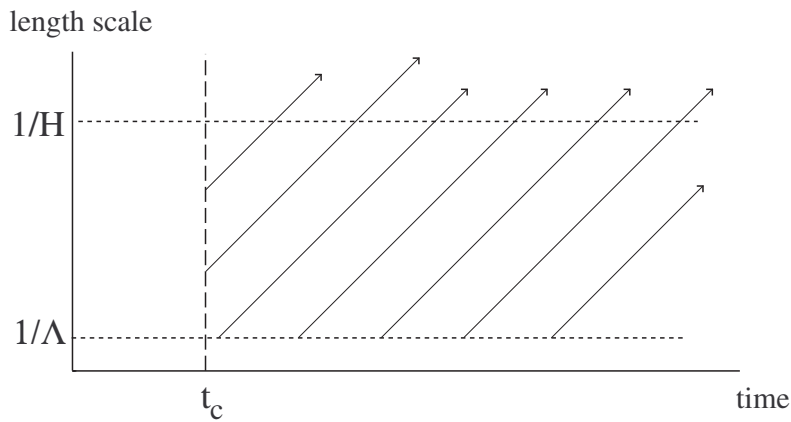}
\caption{The two ``transverse'' modifications of initial conditions during
inflation. The initial conditions are imposed either at a certain scale $%
\Lambda $, or at a certain time $t_{c}$. The modes responsible for structure
formation cross the horizon at $R=1/H$.}
\label{fig:signature}
\end{figure}

In \cite{Kaloper:0307} an analysis very similar to the one in \cite{Ulf:0203}
was performed. The idea was to impose initial conditions different from the
Bunch-Davies vacuum not at a fixed scale at all times, but at a fixed time
on all scales. The main difference is that the fixed scale $\Lambda $ in eq.
(\ref{eq:pk}) should be replaced by the physical scale of the mode at the
initial time. This way of imposing initial conditions breaks de Sitter
invariance and gives rise to a different kind of modulation. There is also a
numerical factor, as explained in \cite{Kaloper:0307}, depending on the
change of the slowroll parameters at the initial time. It is useful to draw
a diagram which illustrates the two possibilities (see Fig. \ref
{fig:signature}).

In \cite{Kaloper:0307} it was argued that one should choose the Bunch-Davies
vacuum at the fundamental scale. As a consequence, as argued in \cite
{Kaloper:0307}, any departure from the Bunch-Davies vacuum will be washed
away on small scales. It is only on the very largest scales that traces of
the initial conditions remain, and if inflation has proceeded longer than
the minimal 60 or 70 e-foldings, the scales where the imprints are located
has not yet reentered (and never will if the universe now starts to
accelerate again).

The aim of the next section is to argue that traces of non-thermal physics
much further back can remain hidden in the trans-Planckian regime and
manifest themselves as modified initial conditions at the fixed scale.

\bigskip

\section{Time scales of thermalization}

\setcounter{equation}{0} \label{sec:therm}

\bigskip

\subsection{Thermalization in de Sitter space}

Having a non-thermal state in a thermal environment suggests that the
construction has a limited lifetime. The question we will try to answer in
this section is how long this lifetime really is. That is, for how long can
inflation, with a non-thermal vacuum, continue before the vacuum relaxes and
becomes thermal? A naive first estimate would suggest that the thermalization
time in de Sitter space would be related to the temperature and be given by $1/T
$, where $T$ is the Gibbons-Hawking temperature \cite{GibbHawk:1977}. It is
easy to see, however, that this crucially depends on the strength of the
interactions that are causing the thermalization. In an expanding universe
the interactions must be faster than the expansion of the universe for the
thermalization to happen in this way. Since the time scale of the expansion
is $H^{-1}\sim T^{-1}$ this depends in a detailed way on the actual
interactions. In a slow-roll inflation, with its flat potential, the
interactions are not fast enough and thermalization will not take place.%
\footnote{%
Note that this assumption is implicit also in \cite{Kaloper:0307} when it is
argued that physics can remember initial conditions at a fixed time all the
way back through many e-foldings.} To summarize, after the fluctuations have
grown larger than the horizon, there is no causal process which could affect
them. As a consequence the fluctuations freeze and any non-thermal signatures
are kept intact.\footnote{One way to see that a thermalization time of order
$1/T \sim R$ is far too short, is to consider our present stage of cosmic acceleration.
If we consider this state as a perturbed de Sitter space, one finds that with a
thermalization time of order $R$ the lifetime for the perturbations
would be roughly $R \sim 10$ billion years. Now our sun for example (being a small
perturbation of the present de Sitter phase) will even $1000$ billion years into
the future still exist as a small perturbation, by then in shape of a cooling white
dwarf.}

This global perspective could, however, be misleading from the point of view
of a local observer \cite{Danielsson:2003cn}. It is important to note that
taking a global perspective of the process ignores possible subtleties
having to do with horizons. Since the physics of horizons, especially in de
Sitter space, is largely mysterious one needs to be careful when they come
into play. After all, the counterpart of black hole complementarity, \cite
{Susskind:1993if,Susskind:1993ki,Susskind:1993mu,
Susskind:1995qc}, is not well understood in de Sitter space (see, e.g.
\cite{Verlinde:0209,UDM:0210}). Let us therefore take a local perspective and try to
figure out what would be observed.

We start with a mode that is created at a definite scale and then expands to
larger size. A local observer would see how the mode approaches the horizon
but never crosses it. The closer the horizon it gets the more redshifted it
becomes. From the local observer's point of view anything approaching the
horizon effectively enters into a region with ever higher temperature and
will at some point in time be completely ``thermalized'' and turn into
radiation. It therefore seems like there is indeed a finite lifetime for any
excitations above the thermal state.

It is important to note that our first argument, where thermalization was
excluded, was made using quantum field theory augmented with Hawking
radiation. It is only when we take a further step, invoking physics related
to the horizon -- eventually described by quantum gravity, holography and
complementarity -- that we are able to find signs of thermalization.

Before presenting the conditions we believe need to be fulfilled in order
to give a correct estimate of the thermalization time, let us comment on
a proposal made in \cite{Kaloper:0209}, where the same issue was discussed.
There thermalization also was invoked based on effects appearing only once
the physics of horizons was taken into account. The condition the authors
of \cite{Kaloper:0209}
claim is determining the thermalization time is the time it takes an object
(initially starting close to the center of the static patch) to be of order 
a Planck length away from the horizon.
Indeed, from the local observers point of view the temperature at that point
is of order the Planck scale, naively implying that gravitational interactions quickly
thermalize the object in question. This estimate of the thermalization
time is easily shown to give $\tau \sim R \ln R$.

We believe this way of arguing suffers from a number of different problems.
First of all, it only concerns objects approaching the observers horizon, it
has nothing to say about the observer herself and things bound to her. In
particular, the time it takes for her to be thermalized,
using this criterion, would be infinite.
This is clearly wrong. In order to overcome this problem one could argue that,
somehow, the time it takes for local objects to thermalize is also of
order $R \ln R$. This, however, must also be wrong, as can be concluded from
footnote 2 (and noting that $\ln R$ only gives a factor of order 100 for our 
present universe).

A second, perhaps more serious, objection concerns the actual validity of
the argument even for objects approaching the horizon. If we focus on what
an observer actually would see as an object recedes towards the horizon, one
needs to be careful. Since the rate of the photons (emerging from the horizon)
received by our observer
is of order $1/R$ (see \cite{UDM:0210}), the time it would take for her to see the object
being thermalized must be a much larger time than $R \ln R$, during which time
only of order $\ln R$ photons could have been detected. In fact, the
time it will take is of order $R^3$, as will be shown below.

The reason we find it important to focus on what actually would be seen is that,
just like in inflation, one can imagine that the de Sitter phase is abruptly
turned off, being followed by a more standard cosmological evolution. When this
happens, the object will, of course, return to the observers causal patch at some
time in the future \cite{UDM:0210}. Since the
situation between the object and the observer is symmetric, it is clear that the
object will be in as good shape as the observer. If the observer has not been
thermalized by then, then neither should the object be, implying that the estimated
thermalization time in \cite{Kaloper:0209} is far too short.

Indeed, considering the symmetric
situation we have between the observer and the object (being for example another 
observer) and the fact that they can meet again some time after the de Sitter
phase has turned off, seems to imply that the estimated thermalization time should
be the same for local objects as for those who approach the horizon, even from the
perspective of one single observer.

With this last observation in mind let us make two independent estimates of the
thermalization time, one considering objects approaching the horizon, and
one considering local object, bound to the observer. We will find that
they are indeed of the same order.

As mentioned above, in order to decide how fast
an object (falling towards the horizon) is being thermalized, one should focus on what
is actually seen by a local observer. From this point of view the thermalization process
will be very slow considering the fact that the rate of photons received by the
observer is of the order $1/R$, i.e. one photon every $R$ \cite{UDM:0210}. Let us now try to
find out what actually happens to the object (according to the observer). To do
that we think of the horizon as an area consisting of $R^2/l_{pl}^{2}$ Planck cells, and
remember that the photon has a wavelength of order Planck scale when emitted
and can indeed resolve specific Planck cells.

Now first assume
the object in question is something really simple, corresponding to an information content
much smaller than the $R^2$ number of degrees of freedom of the horizon. This would mean
that only a few of the Planck cells are involved in encoding the object. In the extreme case
of an object with entropy of $\mathcal{O}(1)$, one would need to wait until of the order $R^2$
photons has been emitted to be reasonably sure to see a photon coming from the burning of
the object. In the other extreme one can think of an object consisting of the order $R^2$ degrees of
freedom. In this case it is clear that one has to wait until of the order
$R^2$ photons has been emitted, in order for all parts of the object to have been
burnt. And so regardless of the size of the object, one has to wait a time,
\begin{equation} \label{eq:thermtime}
\tau \sim \frac{R^2}{l_{pl}^{2}} R = \frac{R^3}{l_{pl}^2},
\end{equation}
in order to actually see the destruction.\footnote{We note that precisely this time scale
was found in \cite{Page:9305}, where in that paper $R^3$ could, in the context of
black holes, be interpreted as the time, after some object
had fallen into the black hole, an observer would need to wait in order to recover the corresponding
information.} At this point one could object and say
that since the blueshift is so strong near the horizon, anything close to it
would thermalize almost immediately, leading to a much shorter thermalization
time. However, taking
the blueshift into account also requires taking the time dilation into account.
Indeed, this makes processes happening close to the horizon appear extremely slow.
This is one way to see that the large blueshift is perfectly consistent
with the relatively long time scale in eq. (\ref{eq:thermtime}).

We therefore suggest that the maximal
lifetime of non-thermal excitations in de Sitter space is given by $\tau \sim
R^{3}/l_{pl}^{2}$. It is important to emphasize that this relatively long
time scale follows only if other non-gravitational interactions are frozen.
If this is not the case, and those interactions occur faster than the
expansion of the universe, the characteristic thermalization time will scale
like the naive $1/T$. In that case we do not need any reference to
holography or complementarity. Thermalization occurs regardless of
perspective.

Now let us try to estimate the time it takes to thermalize a local object,
bound to the observer. Let us reconsider the possibility that local interactions do give rise
to a thermalization, but only if we take physics near the Planck scale into
account. Again it is the Hawking radiation that we expect can do the job.
The probability that the Hawking radiation, with wavelength of order $R$,
will interact (and thereby thermalize) with physics at a fundamental length
scale, say at the Planck length, is small, but obviously nonzero. The
question is, therefore, how long do we need to wait in order for the
probability of scattering of a given Planck cell with the de Sitter
radiation to be of $\mathcal{O}(1)$? The claim is that this will give a
rough estimate of the thermalization time of the underlying microphysics.
The interaction rate is $\Gamma =\sigma nv$, where for this particular
process we have for the cross section $\sigma \sim l_{pl}^{2}$, the number
density of the radiation $n\sim T^{3}\sim 1/R^{3}$ and the relative velocity
$v=c=1$. Then the time $\tau $ it takes for this process to be likely is
given by the condition $\Gamma \tau \sim 1$ implying,
\begin{equation}
1\sim \sigma nv\tau \sim l_{pl}^{2}\cdot 1/R^{3}\cdot 1\cdot \tau \mbox{ }%
\Rightarrow \mbox{ }\tau \sim \frac{R^{3}}{l_{pl}^{2}},
\end{equation}
which is seen to coincide, up to orders of one, with the previous result.
Therefore, regardless of whether local objects or objects falling towards
the horizon are concerned, the thermalization time will be the same. We argued
above that this must be the case based on the symmetry between the observer
and the object and by noting that, if the de Sitter phase is only temporary,
they will meet again. We find it encouraging that the above results are in
agreement with this assessment.

To summarize, we have made three attempts to estimate the thermalization
time. The one where the effects of horizons was ignored
led to an apparent contradiction with an estimate based on horizons and
holography. A reconciliation of these two estimates need an understanding of
complementarity. A reasonable conclusion is, though, that $\tau $ represents
the longest time that a non-thermal deviation can survive from a local
perspective.

We also found that the longer time scale can be obtained from Hawking
radiation acting through Planck scale physics. It would be interesting to
understand this connection better. The Planck scale argument suggests that
the longer time scale could have global importance by affecting the initial
conditions. That is, the vacuum, at Planck scale, would be thermalized after
a time $\tau $.

Now let us see what the implications of these results are for inflation. In
inflation the Hubble constant is constrained from observations to be no
larger than $H\sim 10^{-4}\mbox{ }m_{pl}$. With this input the
thermalization time for non-thermal excitations ($\alpha $-vacua included) is
found to be of order $\tau \sim R^{3}=1/H^{3}\sim 10^{12}\mbox{ }t_{pl}$.
Comparing this with the time needed for the required number of e-foldings,
which for 70 e-foldings is $t_{infl}\sim 70/H\sim 7\cdot 10^{5}\mbox{ }%
t_{pl} $, one can conclude that the thermalization time allows for visible
effects of non-thermal behavior in the CMBR, with room to spare. On the other
hand, if we imagine that inflation went on uninterrupted for a very long
time, thermalization effects become important.\footnote{%
More generally, the thermalization time, in units of e-foldings, relevant
for initial conditions on a scale $\Lambda $ can be expected to be $\left(
\Lambda /H\right) ^{2}$. If the desired amplitude for the corrections to the
primordial spectrum is $10^{-2}$ the number of allowed e-foldings become $%
10^{4}$.}

\bigskip

\subsection{General considerations of thermalization in cosmological like
boxes}

\bigskip

Let us investigate the thermalization a bit further. The time scale is in
general given by
\begin{equation}
\tau \sim \Gamma ^{-1}\sim E^{2}/T^{3}.
\end{equation}
In the expanding de Sitter space, with a constant Hawking temperature, the
expanding modes redshift like $E\sim e^{-Ht}\varepsilon $. Cutting of at the
horizon scale, where everything freezes, we find $E\sim T$ and as a
consequence the naive $\tau \sim 1/T$. A slightly more detailed calculation
focuses on the integral $\int dt \mbox{ } \Gamma $. It is only if this
integral \textit{diverges} before horizon crossing -- i.e. there is time for
an infinite number of interactions -- that we get exact thermalization. In
our case we find
\begin{equation}
\int^{t_{cross}}dt\mbox{ }\Gamma \sim \frac{T^{3}}{H\varepsilon ^{2}}%
(e^{2Ht_{cross}}-1)\sim T/H\sim 1
\end{equation}
where we have used $e^{2Ht_{cross}}\sim \varepsilon H\sim \varepsilon /T\gg
1 $. Note that we have taken a global perspective for our argument. The only
way for the integral to become large is if the interactions are truly
strong, unlike what is the case in inflation, giving rise to a large
dimensionless prefactor in the above expression.

A main point of our paper has been to argue that there could be another way
to get thermalization. We first argued for this from a holographic point of
view. We then found a local derivation which gave the same result. Let us
now proceed along the second line of approach to see whether we can learn
something more. We assume that Planckian physics can be modelled by a fixed $%
E$ of order $\Lambda $ and that the Planckian physics not only affects what
is going on at lower energies, but also suffers a possibly thermalizing back
reaction from low energy physics. Under these assumptions the integral
trivially diverges and the system, as we have argued in the previous
subsection, eventually thermalizes. To be slightly more general, we can
consider a temperature which is not constant (unlike the Gibbons-Hawking
temperature in de Sitter space), but instead redshifts with the expansion of
the universe. We then find
\begin{equation}
\int dt\mbox{ }\Gamma \sim \int dt\mbox{ }\frac{T^{3}}{\Lambda ^{2}}\sim
\int dt\frac{1}{a^{3}\Lambda ^{2}},
\end{equation}
which diverges only if the scale factor grows slower than $t^{1/3}$. For
this to be the case, one needs an equation of state $p=\sigma \rho $ where $%
\sigma >1$ which is unphysical. In conclusion we find that we must put the
system in a box that prevents the temperature from redshifting. This is
precisely what happens in de Sitter space.\footnote{%
It is amusing to speculate whether there are astrophysical systems where the
time scale eq. (\ref{eq:thermtime}) could be important. In this context one
may note that the hottest stars, with internal temperatures of the order 1
GK (or $10^{5}$ eV) give rise to a time scale of $10^{17}$ years, which
clearly is much too long to be of any interest. However, there are several
systems emitting radiation in the TeV range which translated into
temperature becomes $10^{16}$ K. If this is inserted into eq. (\ref
{eq:thermtime}), one finds a relaxation time of about an hour. If our
reasoning is correct, one could speculate that in a region where TeV
temperatures are sustained for an hour or so, effects of quantum gravity
could give rise to a complete thermalization of any physical system
(including baryon number violation etc.).}

\bigskip

\subsection{Remarks on time scales in finite isolated systems}

\bigskip

\begin{figure}[tbh]
\centering
\includegraphics[height=6cm]{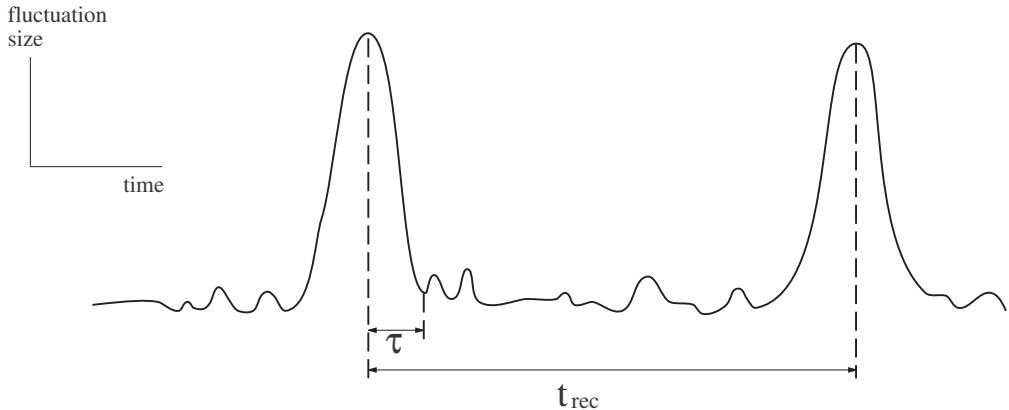}
\caption{ The diagram schematically depicts chance fluctuations of high
entropy in terms of the maximum relaxation time $\protect\tau \sim R^{3}$,
and the Poincar\'{e} time $t_{rec}\sim e^{S}$. }
\label{fig:fluctuation}
\end{figure}

One should note that the time scale we have discussed, although long, is
very much shorter than the Poincar\'{e} recurrence time (see Fig. \ref
{fig:fluctuation}).
The Poincar\'{e} recurrence time is the time scale over which macroscopic
thermal fluctuations are expected to happen (for recent qualitative
discussions on this in the context of de Sitter space, see for example \cite
{UDM:0210,Dyson:0202,Dyson:0208,Banks:0210,Goheer:0212,Kachru:0301}). The
typical time scale is given by $t_{rec}\sim e^{S}$, where $S$ is the entropy
of the system. The second law is only meaningful near a macroscopic
fluctuation and states that time always points towards increasing entropy.
Our time scale eq. (\ref{eq:thermtime}) is the maximum decay time (or, by
time reversal symmetry, rise time) of a macroscopic fluctuation.\footnote{%
In \cite{UDM:0210} it was noted that there is an interesting geometrical
connection between these time scales in a space time where inflation
suddenly stops.}

We suggest that this is a general behavior for finite isolated quantum
systems, not limited to de Sitter space. Another such system of particular
interest is large black holes in Anti de Sitter space (for discussions on
long, and very long, time scales in this context, see for example \cite
{Maldacena:0106,Kraus,Fidkowski,Barbon}). Since these black holes are
eternal they are particularly intriguing laboratories to test some of the
ideas mentioned above. Interestingly, both time scales (maximal relaxation
and recurrence) seems to tell us something about retrieval of information
that has fallen into the black hole \cite
{Maldacena:0106,Kraus,Fidkowski,Barbon}. The mechanism for this and the
meaning of these time scales is, however, still obscure \cite{Barbon}. Even
though tempting, we do not to attempt to address these issues here.\footnote{%
It is important to point out though that in the discussion of thermalization
in de Sitter space, we mean thermalization strictly in the thermodynamic
sense. We do not mean actual loss of coherence at the microscopic level.
However, these residual correlations are not likely to be detectable and so
effectively the spectrum would be thermal.}

\bigskip

\section{Conclusions}

\setcounter{equation}{0} \label{sec:concl}

\bigskip

In this paper we have investigated the problem of thermalization in de
Sitter space. We have argued, based on holography, that the maximum
thermalization time -- relevant if all non-gravitational interactions are
weak and slower than the expansion of the universe -- is given by eq. (\ref
{eq:thermtime}).\footnote{%
This is provided that the trans-Planckian physics can be viewed as a closed
system and there are no sources of negentropy.} We have also argued that
this allows trans-Planckian physics to retain a memory of physics preceding
inflation for up to $N=H\cdot \tau \sim 1/R\cdot R^{3}=R^{2}/l_{pl}^{2}$ e-foldings,
which for $R\sim 10^{4}\mbox{ }l_{pl}$ gives $N\sim 10^{8}$.

We have also speculated that the time scale could have a more universal
importance and not only be relevant for de Sitter space. In particular there
seems to be two typical time scales involved in box-like configurations. One
is the maximal relaxation time (which we interpreted in terms of
thermalization in de Sitter space), which is the typical decay (or rise)
time for macroscopic fluctuations. The other is the recurrence time, which
is the typical time scale for which these macroscopic fluctuations are
expected to happen.

\section{Acknowledgments}

UD is a Royal Swedish Academy of Sciences Research Fellow supported by a
grant from the Knut and Alice Wallenberg Foundation. The work was also
supported by the Swedish Research Council (VR).

\end{document}